\documentstyle[11pt]{article}

\begin{document}
\begin{flushright}
GUTPA/99/05/2\\
\end{flushright}
\vskip .1in
\newcommand{\lapprox}{\raisebox{-0.5ex}{$\
\stackrel{\textstyle<}{\textstyle\sim}\ $}}
\newcommand{\gapprox}{\raisebox{-0.5ex}{$\
\stackrel{\textstyle>}{\textstyle\sim}\ $}}
\newcommand{\lsim}{\raisebox{-0.5ex}{$\
\stackrel{\textstyle<}{\textstyle\sim}\ $}}

\begin{center}

{\Large \bf  Why do we have parity violation?}

\vspace{20pt}

{\bf C.D. Froggatt}

\vspace{6pt}

{ \em Department of Physics and Astronomy\\
 Glasgow University, Glasgow G12 8QQ,
Scotland\\}

\vspace{6pt}

and \\

\vspace{6pt}

{\bf H. B. Nielsen} \\

\vspace{6pt}

{ \em Niels Bohr Institute\\
 Blegdamsvej 17, Copenhagen $\phi$,
Denmark}

\end{center}

\section*{ }
\begin{center}
{\large\bf Abstract}
\end{center}

We discuss here two of the questions posed at the beginning of the Bled 1998
workshop:
Why is the weak charge dependent on handedness?
Why do we have parity violation in the Standard Model?
It is argued
that the quarks and leptons must be protected from
gaining a fundamental mass, very large compared to the
electroweak scale, by gauge invariance and hence that their gauge charges
must depend on handedness. Furthermore we argue that it is the conservation
of parity in the electromagnetic and strong interactions rather than
parity violation in the weak interactions that needs an explanation.
We derive this parity conservation and indeed the whole system of
Weyl fermion representations in the Standard Model from a few simple
assumptions:
Mass protection, small representations, anomaly cancellation and
the Standard Model gauge group $S(U(2)\times U(3))$.

\vspace{80pt}

Updated version of discussion published in the
Proceedings of the International Workshop on
{\it What comes beyond the Standard Model}, Bled, Slovenia,
29 June - 9 July 1998 (DMFA - zalo\u{z}ni\u{s}tvo, Ljubljana).

\thispagestyle{empty}
\newpage

\section{Introduction}

Why do we have parity violation, or why is the weak charge dependent
on handedness?

The short answer to this question is that
we need at least some of the charges to be different for
the observed right-handed and left-handed fermion
states---i.e. handedness dependent or chiral---for the
purpose of {\em mass protection}.
That is to say our philosophy is that the particles we ``see''---those
we can afford to produce and measure---are very
light (essentially massless) from the supposed
fundamental (GUT, Planck,...) scale point of view.
Consequently they need a mechanism for being
exceptionally light, so that we have a chance to ``see' them.
This mass protection mechanism is suggestively provided \cite{blcdf}
by requiring  that any pair of right and
left (Weyl) components, for the observed quarks and leptons,
should have at least one gauge quantum number different between them,
so that any mass term is forbidden by gauge invariance.

Really we should rather ask why is parity conserved in the
electromagnetic and strong interactions. Our philosophy would be that
{\em a priori} there is no reason why these symmetries should be
there at all, and it is the presence of the
symmetries (rather than their breaking)
that needs an explanation \cite{OrigOfSym}. This is the philosophy
of what we call random dynamics, which really means:
all that is not forbidden occurs. It is a very natural
assumption, since really
to know a symmetry exists is much more informative
than to know it not to be there.
So {\em a priori} one should rather say that,
if there is no reason for them,
we should not expect symmetries to be present.

In the case of the question of
whether the electroweak charges on the Weyl components of the quark and
charged lepton fields should be the same for
the two handednesses---right and left---we can say
that, since the Weyl fields transform
under Lorentz transformations without mixing into
each other
(i.e.~they transform into themselves only),
we should consider each
Weyl field as essentially corresponding to a completely separate
particle. As separate particles we expect them to {\em a priori} have
completely different charges.
You might of course object that when the particle has a mass, so that we
are talking about a Dirac particle, there is a connection between
the left and the right Weyl component fields.
However in the Standard Model
it is well-known that the masses come about as an effect of the Higgs
field vacuum expectation value. So, before the
effect of the Higgs mechanism,
the fermions are massless and there is no association of the various Weyl
fields with each other {\em a priori}.
It is therefore not surprising that the weak charge is dependent
on handedness and, as emphasized above, it can then provide
a mass protection mechanism.

The question that deserves and needs an answer is rather
why there is parity conservation for the strong and
elctromagnetic interactions,
in the sense that the electromagnetic and colour charges are the same
on the right and the left components of the same Dirac particle.
We show, in section 3, that this result follows from the requirements of
small representations for the strong and electromagnetic interaction
gauge {\em group} $U(3)$ \cite{ORaif} and of
no gauge anomaly for the photon and gluons.
In addition the fact that the right-handed components are singlets,
while the left-handed components are doublets, under the weak
SU(2) gauge group needs an explanation.
In section 4, we show that this result can also be derived, by extending
the above requirements to the full Standard Model gauge
{\em group}
$S(U(2)\otimes U(3))$ \cite{ORaif,blgroup} and
also requiring the quarks and leptons to be mass protected.
Indeed we derive the complete representation pattern of the
Standard Model fermions. We present our conclusions in section 5.

So let us begin by stating our basic assumptions about the full
Standard Model and then specialise these to the assumptions necessary
to derive parity conservation in the strong and electromagnetic
interactions.


\section{Starting assumptions}
\subsection{The assumptions to derive Standard Model representations of
fermions \label{SMassum}}

(a) As the starting point for the derivation of the Standard Model
representations, we
shall assume
the gauge \underline{group} and not only the gauge Lie algebra
of the Standard Model to be $S(U(2)\otimes(U(3))$.

\noindent (b) Further we shall make the assumption that the
representations---realised by the
Weyl fermions---of this group are ``small''. More specifically
we assume
that the weak hypercharge charge $y/2$ is
at most unity numerically, and that only the trivial and the lowest
fundamental (defining) representations of the nonabelian groups SU(2) and
SU(3) are used.

\noindent (c) Further  we assume mass
protection, i.e.~we say that all particles
for which a mass could be made, without the Higgs field being used,
would be so heavy that we should not count them as observable particles.

\noindent (d) In our argument we shall also use
the requirement that there shall
be no gauge nor mixed anomalies. This is needed
since otherwise there would be a breaking of the gauge symmetry.

These assumptions are of course known to be true in the Standard Model.
Indeed they are rather suggestive regularities of the Standard Model,
if one is looking for inspiration to go beyond the Standard Model.
You could say that it might not be so difficult to find some rather
general argumentation for why representations should be ``small''
in some way---not exactly how small perhaps.

We made a similar set of assumptions in
the appendix of ref.~\cite{fns} to derive
the representations in a Standard Model generation of fermions.
The assumption that the weak hypercharge $y/2$ is at most unity
numerically was replaced there by the assumption that the sum of weak
hypercharge squared for all fermions in a generation is as small
as possible. Related discussions can be found
in refs.~\cite{Marshak,Volkas}.

\subsection{Slightly reduced assumptions for parity in strong and
electromagnetic interactions}

{}From the
assumptions stated in the foregoing subsection we can indeed
derive the fermion representations of the whole Standard Model
and, thus, also the fact that there is parity conservation in
electromagnetic and strong interactions.
However, if we replace the requirement $|y/2|\leq 1 $ by the
slightly modified assumption
that the electric charge $Q=y/2+I_{W3}$  (where $I_{W3}$
is the third component of the weak isospin)
has numerical value less than
or equal to unity for all the Weyl fermion representations,
the mass protection assumption is not needed
for this parity derivation. The point of course is
that the mass protection
is performed by the weak interaction, and
the electromagnetic and colour quantum numbers do not provide any
mass protection themselves---they cannot with parity symmetry.

In other words, for the derivation of
parity conservation in strong and electromagnetic interactions
alone, we assume:

(a) Either the gauge \underline{group} $U(3)$ for strong and
electromagnetic interactions, or the total
gauge group $S(U(2)\otimes U(3))$ as above.

(b) The ``small'' representations in the form $|Q|\leq 1$ and
$|\underline{a}| \leq |\underline{3}|$.

(c) Then of course there should still be no anomalies.

\section{Derivation of parity for QCD and electromagnetism}

The program of our proof of parity conservation
for strong and electromagnetic
interactions consists in showing that
the Weyl fields must have quantum number combinations
that will be paired into
Dirac fields, so that parity in the electromagnetic and strong interactions
gets preserved.

We should of course have in mind that, in
four dimensions, one can consider the right-handed Weyl components
as represented by their CP-conjugates
so to speak, meaning a corresponding
set of left-handed fields with the opposite charges.
So we actually
need only discuss the
left-handed components, just letting them represent the right-handed
ones too as antiparticles.
What we then have to show is that there are
always equally many left-handed Weyl field species
with a given electric and colour charge combination and the opposite.
In this way we could then say that at least the possibility is there for
combining these Weyl fields into Dirac fields, so that the
electric and colour charges on the right and the left Weyl components
become the same.

Now, for anomaly calculations, it is easily seen that
left-handed Weyl fields in conjugate
representations give just equal and
opposite contributions to the various
anomalies. Thus we can only hope to say from anomaly considerations
something about the number of species in one representation minus
the number in the conjugate one. We should therefore introduce names
for these differences:

We let the symbol $ N_{(y/2, \underline{a}, I_W )}$
denote the number of
left-handed Weyl species with the weak hypercharge $y/2$, the colour
representation $\underline{a}$ and the weak isospin $I_W$ minus
the number of species with the opposite (conjugate)  quantum numbers.
But, in the present section, we
ignore the weak isospin and use  $N_{(Q , \underline{a})}$ to mean
the difference between the number of Weyl-species with electric charge
Q and colour representation $\underline{a}$ and the number of Weyl-species
with the conjugate quantum numbers.

The requirement of the smallness of the representations
means that  $ N_{(Q, \underline{a})}$ is zero unless
\begin{equation}
|Q|\leq 1
\end{equation}
\begin{equation}
|\underline{a}|\leq |\underline{3}|
\end{equation}
Obviously by our definition $N_{(Q =0,\underline{1})} = n - n =0$,
since the representation $(Q =0,\underline{1})$ is self-conjugate.
The requirement of the gauge \underline{group}
being $U(3)$ means that the representations must satisfy the
electric charge quantisation rule \cite{blgroup}
\begin{equation}
Q + t/3 = 0 \qquad (\mbox{mod 1})
\end{equation}
where $t$ is the triality of a representation.

What we have to show to get parity conservation for these
interactions is that
\begin{equation}
N_{(Q,\underline{a})}=0
\end{equation}
for all the quantum number combinations $(Q, \underline{a})$.

The requirements of small representations
and of the gauge \underline{group} being
$U(3)$ leaves only the three differences of species numbers
$ N_{(Q=1,\underline{1})}$, $ N_{(Q=2/3 ,\underline{3})}$,
$ N_{(Q=-1/3,\underline{3})}$ non-zero.

Now the anomalies in four dimensions come from triangle diagrams
with external gauge fields for the gauge anomalies and with two gravitons
and one gauge particle assigned in the case of the mixed anomaly.
In order to get rid of the anomalies, so as to avoid
breaking the gauge symmetry say,
we must require that the relevant triangle diagrams
have cancellations between the contributions coming from the different
Weyl field species that can circle around the triangle.
The only mixed anomaly diagram, not already vanishing for other reasons,
is a triangle with Weyl particles circling around it having
two gravitons attached and the photon
at the third vertex. The cancellation required to get rid of this
the mixed anomaly becomes
\begin{equation}
N_{(Q=1,\underline{1})}+\frac{2}{3} \times 3N_{(Q=2/3,\underline{3})}
+(-\frac{1}{3}) \times 3N_{(Q=-1/3,\underline{3})} =0
\end{equation}
To ensure no gauge anomaly there are three triangle diagrams that must
have a cancellation: one with three external gluons, which gives
\begin{equation}
N_{(Q=2/3,\underline{3})}
+N_{(Q=-1/3,\underline{3})} =0,
\end{equation}
one with one photon and two gluons attached, which gives
\begin{equation}
2/3 \times N_{(Q=2/3,\underline{3})}
+(-1/3) \times N_{(Q=-1/3,\underline{3})} =0
\end{equation}
and finally one with three photons attached, which gives
\begin{equation}
N_{(Q=1,\underline{1})}+\left(\frac{2}{3}\right)^3
\times 3N_{(Q=2/3,\underline{3})}
+\left(-\frac{1}{3}\right)^3 \times 3N_{(Q=-1/3,\underline{3})} =0
\end{equation}

We have here got four linear equations for three unknowns,
so it is no wonder that they lead to all the differences
$N_{(Q,\underline{a})}$ being
zero. That then means to every Weyl representation there is
the possibility of finding just one with the opposite (conjugate)
representation. This vanishing of the differences $N_{(Q,\underline{a})}$
shows that the gauge theory of colour and electromagnetism is
vectorlike and thus automatically parity conserving,
provided possible mass generation
mechanisms do not violate the gauge symmetries.
It means that one may directly construct
a parity operator, by diagonalizing a perhaps present U(3) gauge invariant
mass matrix and letting it map the right-handed to the corresponding
left-handed
mass eigenstate and vice versa.

\section{Deriving all standard model fermion representations}

Using a very similar technique, but now within all the four assumptions
stated in the subsection \ref{SMassum}, we can show the fermion
representations to be those of the Standard Model with some as
yet not determined number of generations.

For this purpose the assumption about small representations can be taken
to mean
that  $ N_{(y/2, \underline{a}, I_W )}$ is zero unless
\begin{equation}
|y/2|\leq 1
\label{e1}
\end{equation}
\begin{equation}
|\underline{a}|\leq |\underline{3}|
\label{e2}
\end{equation}
\begin{equation}
|I_W|\leq 1/2.
\label{e3}
\end{equation}
Really it means that we assume zero species for the representations
not fullfilling this and thus, of course, the same for the differences
$N_{(y/2, \underline{a}, I_W)}$.
The requirement of the gauge \underline{group} being
$S(U(2)\times U(3))$ means that the species numbers are zero unless
the congruence
\begin{equation}
y/2 + t/3 + d/2 = 0 \qquad (\mbox{mod 1})
\end{equation}
is fullfilled \cite{blgroup}, where $t$ is triality and $d$ is duality.

Obviously by our definition $N_{(y/2 =0,\underline{1}, I_W=0)} =n - n=0$,
since the representation $(y/2 =0,\underline{1}, I_W=0)$
is self-conjugate.

The small representation and the gauge \underline{group} requirements
now allow six $N_{(y/2,\underline{a}, I_W )}$'s to be nonzero
{\em a priori},
namely one for each of the allowed numerical values of $y/2$ which run
from 1/6, in steps of 1/6, to 1.

As above we use the cancellation criteria for the anomalies, meaning the
cancellation of triangle diagrams: This time the mixed anomaly cancellation
diagram has two gravitons and one weak hypercharge coupling and it gives
\begin{eqnarray}
\frac{1}{6} \times 6N_{(y/2 =1/6,\underline{3},I_W=1/2)}+
\frac{1}{3} \times 3N_{(y/2 = 1/3,\underline{\bar{3}},
I_W=0)} & & \nonumber \\
+\frac{1}{2} \times 2N_{(y/2 = 1/2,\underline{1},I_W=1/2)}
+ \frac{2}{3} \times 3N_{(y/2= 2/3,\underline{3},I_W=0)}
& & \nonumber \\
+ \frac{5}{6} \times 6N_{(y/2 =5/6,
\underline{\bar{3}},I_W=1/2)}+N_{(y/2=1,\underline{1},I_W=0)} & = & 0
\end{eqnarray}
The no gauge anomaly triangle diagrams consist of one with three external
gluons, while the one with three external W's
is trivially zero and does not count. Then there are two diagrams with
two gluons and one weak hypercharge coupling and
two W's and one weak hypercharge coupling respectively.
Finally there is one diagram with all three attached gauge particles being
the abelian one (coupling to weak hypercharge). The
corresponding anomaly cancellation conditions become:
\begin{eqnarray}
2N_{(y/2 =1/6,\underline{3},I_W=1/2)}
-N_{(y/2 = 1/3,\underline{\bar{3}},
I_W=0)} & & \nonumber \\
+ N_{(y/2= 2/3,\underline{3},I_W=0)}-2N_{(y/2 =5/6,
\underline{\bar{3}},I_W=1/2)} & =& 0,
\end{eqnarray}
\begin{eqnarray}
\frac{1}{6} \times 2N_{(y/2 =1/6,\underline{3},I_W=1/2)}+
\frac{1}{3} \times N_{(y/2 = 1/3,\underline{\bar{3}},
I_W=0)} & & \nonumber \\
+ \frac{2}{3} \times N_{(y/2= 2/3,\underline{3},I_W=0)}
+\frac{5}{6} \times 2N_{(y/2 =5/6,
\underline{\bar{3}},I_W=1/2)} & = & 0,
\end{eqnarray}
\begin{eqnarray}
\frac{1}{6} \times 3N_{(y/2 =1/6,\underline{3},I_W=1/2)}
+\frac{1}{2} \times N_{(y/2 = 1/2,\underline{1},I_W=1/2)}
& & \nonumber \\
+ \frac{5}{6} \times 3N_{(y/2 =5/6,
\underline{\bar{3}},I_W=1/2)} & = & 0
\end{eqnarray}
and
\begin{eqnarray}
\left(\frac{1}{6}\right)^3 \times 6N_{(y/2 =1/6,\underline{3},I_W=1/2)}+
\left(\frac{1}{3}\right)^3 \times 3N_{(y/2 = 1/3,\underline{\bar{3}},
I_W=0)} & & \nonumber \\
+\left(\frac{1}{2}\right)^3 \times 2N_{(y/2 = 1/2,\underline{1},I_W=1/2)}+
\left(\frac{2}{3}\right)^3 \times 3N_{(y/2= 2/3,\underline{3},I_W=0)}
& & \nonumber \\
+\left(\frac{5}{6}\right)^3 \times 6N_{(y/2 =5/6,
\underline{\bar{3}},I_W=1/2)}+N_{(y/2=1,\underline{1},I_W=0)}
& = & 0
\end{eqnarray}

Here we have got 5 equations linear in the $N$'s of which there are
6. Thus it is not surprising that there is, up to the unavoidable
scaling by a common factor of all the unknowns---the N's, just
one solution. This must, however, be that of the Standard Model,
since the latter satisfies the anomaly cancellation conditions.
The scaling factor corresponds to the
generation number we could say.
So far we have only shown that the N's
are as in the Standard Model. We now need to use the assumption
about mass protection to deduce that we cannot have both
representations---i.e.~a representation and its
conjugate---associated with a given N
present. That implies first that the cases of N's that are zero
imply that there will be no Weyl fermions at all associated with those
quantum numbers---there will be no vector fermions. Also for the cases
of nonzero N's only one of the two associated representations will
exist, depending on the sign of the N in question.
With this conclusion we almost truly derived the Standard Model
fermion representations. There are however still two ambiguities:
1) the generation number can be any integer, 2) we could have the
opposite signs for the N's which would correspond to a model
that is, so to speak, a parity reflected version of the Standard Model.

Since we have now derived the whole representation system for the
fermions in the Standard Model, we did not really need the exercise
of deriving parity for the electromagnetic and colour interactions
separately; we got it all at once after all, assuming though---as is
needed---the Higgs mechanism for mass generation.

\section{Conclusion}
We have shown that, from the four requirements listed in subsection 2.1,
it is possible to argue for the whole system of Weyl fermion
representations in the Standard Model.
So if one can just argue for these assumptions in some model beyond the
Standard Model, one will have the fermion system for free.

Concerning the question of whether the charges depend on handedness,
we saw that for the colour and electric charges no such dependence is
allowed, by just using the smallness of electric charge and
colour representation plus the no anomaly conditions.
Concerning the question of why there is a dependence---namely for the
weak charges---we saw that it was the mass protection requirement that
enforced it. In fact each of the differences N had to be
a difference between zero and another number, because the mass protection
would not allow two sets of Weyl fields counted as left-handed having
opposite (conjugate) quantum numbers. They would namely combine
to get a huge mass and be unobservable, according to
the philosophy of mass protection.
Thus indeed the charges must, in one way or another, be different
for the right and left handnesses.

This
really means that we take the point of view that the fundamental
scale, or the next level in fundamentality,
has so huge a characteristic energy
or mass scale that all the particles we know
must, in first approximation,
be arranged to be massless, i.e.~they must be mass protected.

\section{Acknowledgement}

Financial support of INTAS grant INTAS-93-3316-ext
is gratefully acknowledged.

\end{document}